\documentclass[aps,prd,preprint,groupedaddress,nofootinbib,byrevtex]{revtex4-1}
\usepackage{amsmath,amssymb}
\usepackage[dvips]{graphicx}
\usepackage{mathrsfs}
\begin{document}
\title{  On Experimental Confirmation of the Corrections to the Fermi's golden rule}
\author{Kenzo Ishikawa${}^{1,2}$} \author{Osamu Jinnouch${}^3$}\author{Arisa Kubota${}^3$ }\author{Terrey  Sloan ${}^{4}$} \author{Takuya H. Tatsuishi${}^1$}\author{Risa  Ushiosa${}^3$ }
\affiliation{${}^1$Department of Physics, Faculty of Science, Hokkaido
University, Sapporo 060-0810, Japan}  
 \affiliation{${}^2$   Natural Science center , Keio University , Hiyoshi Japan} 
     \affiliation{${}^3$Department of Physics, Faculty of Science \\
Tokyo Institute of Technology, Tokyo  , Japan}
\affiliation{${}^4$Department of Physics, Faculty of Science \\
Lancaster University, Lancaster ,  UK}
\date{\today}

\begin{abstract}%
Standard calculations by the Fermi's Golden Rule involve approximations. These approximations could lead to deviations from the predictions of the standard model as discused in another paper. In this paper we propose experimental searches for such deviations in the two photon spectra from the decay of the neutral pion in the process $\phi \rightarrow \pi^+ \pi^- \pi^0$ and in the annihilation of the positron from nuclear $\beta$ decay.
\end{abstract}


\maketitle

\section{The correction to the Fermi's golden rule }

In interacting many-body   system described by  a Hamiltonian $H_0+H_1$, a state evolves with  a Schr\"{o}dinger   equation. One particle state is specified  by the momentum and  non-interacting  energy defined by $H_0$. A transition by $H_1$
  has been studied by the Fermi's golden rule. Although these   transitions  have been paid attention from researchers, those that  do not conserve the  energy often arise,   when the approximations  are taken into account.  The   Schr\"{o}dinger equation includes these approximations, which  affect transitions of any states.  Surprisingly,  a correction term beyond  the Fermi's golden rule emerges.  The correction becomes manifest in a transition of a finite time interval, in which that  reveals  different dependence on the 
 time interval  and on the energy difference.  The correction terms would have been identified from experimental data.  However it is not simple as was naively thought due to several reasons.
One reason is that  those  signals that are caused by the corrections terms are similar to those of experimental background. In majority cases, they were considered
 as the background, and  discarded.
 Another reason is on a difficulty to find the absolute value of the physical quantity in experiments, because the data is always modified by an efficiency of detector.  
   The transition rate  describes average behavior \cite{Dirac, Schiff-golden} of the process, and   
  the correction terms give dominant contribution to rapidly changing processes.  Direct observation of these events might give signals of the correction terms, but 
  has not been possible up to the present.   In general it  is difficult to separate  these  from the real background.

Accordingly the correction terms  was not a major concern from  researchers.  Nevertheless,   the correction is one part of the total probability and contributes to  natural phenomena.  Fitting these experiments in approximate  way without the correction term   might be 
possible and viable for certain period.       However, that  should lead serious inconsistency or fatal outcome at later time, which must be avoided.  
It is urgent to confirm an existence of the correction term with simple and clean experiments.

    Two photon processes of  the neutral pion and the positron annihilation supply precise information on the transitions and can be candidates.  
         The rates      have theoretically been well-understood,    and    determined  
from the various experiments, in which the background have  been subtracted.  There are subtlety on  the  background subtraction, and a signal of the correction term has been insignificant.  
The correction terms   are  computed  in a separate paper \cite{ishikawa-oda-nakatsuka}and are found sizable.  Due to their unusual  properties, which  will be presented later,  it is not
 an easy task to disentangle them   from the real background.  Nevertheless,  they 
give universal   contributions   to the phenomena. It will be shown that these are  feasible in $\phi$-factory for the pion and  in nuclear beta decays for the positron.   

   The  neutral pion, $\pi^0$, is  the lightest hadron composed of the quark and anti-quark and  supplies  many informations on particle physics \cite{particle-data}.
 The rate of $\pi^0$ decay to two photons \cite{Fukuda-Miyamoto, Steinberger}   is proportional to the number of the color  $N_c$ \cite{Adler,Jackiw-Bell}, and  the  measurement on life time  $\tau=10^{-16} $ seconds determines  $N_c=3$.  Despite of this remarkable  success, the average life-time  obtained from 
 various methods \cite{Bernstein}  has  large uncertainty of about $10$ per cent. Accordingly, K-meson decays to two or three pions  have also large uncertainties \cite{particle-data}.
 A large uncertainty arises also in the decay of para-positronium, which  is a bound state of the electron and positron
in  Quantum Electrodynamics (QED).   Its properties and transition rates are understood well, but the precision is not very good.   
The large uncertainty of the experimental values may suggest   a fundamental problem on  the transition probability.   
 
 We find the  many body wave function $|\Psi
 \rangle$ composed of normalized states  from   the  Schr\"{o}dinger equation   
 \begin{eqnarray}
i\hbar\frac{\partial}{  \partial t}|\Psi\rangle=(H_0+H_{\text {int}})|\Psi \rangle,
\label{Schroedinger}
\end{eqnarray} 
 , where   $H_0$ and $H_\text{int}$ are the free and interaction parts, and  compute the rigorous transition amplitude.
Hereafter we employ the natural units $\hbar=c=1$ unless otherwise stated.
A transition probability from a state $| i ,0\rangle$ at $t=0$ to a state $|f, T \rangle $ at $t=T$ is  determined  by the von Neumann's fundamental principle of the quantum mechanics (FQM) as , $ P(T)=|\langle f ,T | i,0  \rangle |^2$, for  normalized states.  For $ P(T) \ll 1$, 
   the average rate $\Gamma=\frac{ P(T)-P(T_i)}{T-T_i}$ between  a small $ T_i$ and a large $T$, is given       from a ratio of fluxes of  out-going waves over that of  incident waves and is in agreement with that  derived from the golden rule for the final state of continuous spectrum.  
     In these standard calculations,   the plane waves and the interaction that switches off adiabatically (ASI) are used.  Although, this value  has been used in the majority of the processes,   
       experiments are made at the finite time intervals and the value is measured  
     without average.

      Theoretical values under these conditions are necessary. 
     
           Stueckelberg studied this problem sometime ago and found that the  transition amplitudes of the plane waves  for finite-time interval lead a divergence  \cite{Stueckelberg} even in the tree level. This is unconnected with the ultraviolet divergences due to the intermediate states but to non-normalized initial and final states.  It is possible to avoid this difficulty by using the normalized states.  Those   computed  in the previous paper \cite{ishikawa-oda-nakatsuka} are applied to  experiments.

 $ P(T)$ at a large $T$ is the sum,   
\begin{align}
 P(T)
	&=	\Gamma  T+ P^{(d)} , &
P^{(d)}
	&=	 P(T_i)-\Gamma T_i. \label{probability1}
\end{align}
  $T_i$  is determined by a time that the initial wave packets separates. This is determined by  $\sqrt \sigma_{i}$, where $\sigma_{i}$ is the spatial size 
  of the initial  wave.  At     $\Gamma T_i <1 $, and at $T >T_i$, $ P^{(d)}$ is constant.

  $\Gamma $,   is computed  with the standard S-matrix   $S[\infty]$ under ASI \cite{Goldberger,newton,taylor}, but $ P^{(d)}$ is computed  with  the
   wavefunctions 
        following FQM  \cite{ishikawa-shimomura-PTP, ishikawa-tobita-PTEP,ishikawa-tobita-ANP,ishikawa-tajima-tobita-PTEP}. 
     A  rigorous probability will be obtained without facing the difficulty raised 
     by Stueckelberg by using  the wave packets instead of the plane waves.

Experimental proof of  $P^{(d)}$ in the neutral pion decay,  the positronium decay,  and the positron annihilations  are studied.  Two photon decays of a para-positronium is  almost equivalent to the neutral pion decay. Their systematic analyses are presented.     
It will be shown  that the unique properties derived from  the probability  $\Gamma T+ P^{(d)}$ can be confirmed experimentally. 
  
  The paper is organized as follows: In Section 2, the   pion decay is analyzed and in Section 3, the positron annihilation is analyzed.  In Section 4 the wave packets sizes and relevant parameters are estimated.   In Section 5 the experiments are studied  and summary and prospects are presented.  Appendix A is devoted to various formula and Appendix B is devoted to a method for entanglement of the accidental background.

\section{ Two photon decay of
 the neutral pion  }
 The  interaction of a neutral pion or a para-positronium with  two photons  are
   derived from the triangle diagram of the quark or the electron as
$L_\text{int}
	=	-g\,\varphi\,\epsilon_{\mu\nu\rho\sigma} F^{\mu\nu} F^{\rho\sigma}$
in which the coupling for the pion is $g=\frac{\alpha}{4\pi f_{\pi}}$
 is almost constant from the confining mechanism and   is related with the $\pi\gamma\gamma$ coupling~\cite{Adler,Jackiw-Bell}.
For the positronium, the binding energy is small and the coupling varies with the momentum, which will be ignored for a while.  Substituting this   to Eq.$(\ref{Schroedinger} )$, we have the transition amplitude  for an initial state of a central  momentum and position 
into two photons  
\begin{eqnarray}
 \mathcal{M}=\int_{ T_{\pi}}^{T_{\gamma}}  dt  \int d^3x  
 \langle { K}_1,X_1;{ K}_2,X_2
 |H_\text{int}(x)| { P}_{\pi} , X_{\pi} \rangle 
\end{eqnarray}
 Gaussian wave packet~\cite{ishikawa-shimomura-PTP} ,\cite{peierls},\cite{LSZ} ,\cite{Low} satisfies the minimum uncertainty,  which is  idealistic  for studying  a transition of  finite-time interval, and  is   used in majority of places.  
  Non-Gaussian form is also physically relevant and studied later.  Wave packets of the size $\sigma_i$,  the central momentum, and the central position are used throughout this paper, 
 where $i=1,2$, $E_A=\sqrt{P_A^2+m_A^2},E({ K}_i)=| K_i|$, and
${\vec V}_A = \frac {{\vec P}_A}{ E_A}$
is the group velocity  of the momentum~$\vec P_A$.
Throughout this paper, the upper-case roman letters $A,B,\dots$ run for $\pi,1,2$ so that e.g.\ $\sum_A$ stands for $\sum_{A=\pi,1,2}$, etc.
An imaginary part is added to the energy of  the unstable initial state   according to Ref.~\cite{Weisskopf:1930au,Weisskopf:1930ps}; see also e.g.\ Ref.~\cite{Goldberger} for a review is taken.
Integration over the space position leads to a  Gaussian function in the momentum difference, and that over the time leads  to 
\footnote{
We have put the central momentum $K_i$ in the polarization $\epsilon^\mu$ and in the derivative interaction.
See Ref.~\cite{ishikawa-tajima-tobita-PTEP} for its justification.}
\begin{equation}
\mathcal{M}
	=	N_0({\vec X}_i)
		\epsilon_{\mu\nu\rho\sigma}\epsilon^{\mu}(K_1) K_1^{\nu}\epsilon^{\rho}(K_2) K_2^{\sigma}
e^{-{\sigma_s\over 2}({\delta P})^2}\,G({\delta\omega}),\label{total-amplitude_{pi}}
\end{equation}
where  $N_0({\vec X}_i)$ shows a dependence  on the positions, $\delta { P} 
	={ P}_{\pi}-{ K}_1-{ K}_2, \delta \omega= E_{\pi}-E_1-E_2-{\vec V_0} {\delta { P}} $ ,  and $\sigma_s=\left( \sum_A \frac{1}{\sigma_A}\right)^{-1}$, $  G({\delta \omega})$   is expressed with the error function erf(x+iy). Their  explicit forms are given in \cite{ishikawa-oda-nakatsuka}.
	The transition probability is written as,
\begin{equation}
 P= {1\over 2}\int d^3X_1 d^3X_2 {d^3K_1 \over (2\pi)^3} 
  \frac{ d^3K_2}{(2\pi)^3} \left|N_0 ({\vec X}_i)\right|^2 2\left(K_1 \cdot K_2\right)^2  e^{    -{\sigma_s} ({\delta P})^2}\left|G({\delta \omega})\right|^2,
\label{total-probability_{pi}}
\end{equation}

  As is shown in Ref.~\cite{ishikawa-oda-nakatsuka}  in details ,  $ G({\delta \omega})$   depends on  an intersection of the trajectories determined by the positions of ${\vec X}_i;i=1,2$. If they intersect outside of the material, the interaction does not occur and the amplitude vanishes. If that is inside of the material,  the interaction occurs. This is a bulk region. In the boundary region, the interaction occurs partly. This is the boundary region.

The  integration  in the bulk is proportional to the time interval  due to  the translational invariance  along 
 the initial momentum, and that in the  boundary   is proportional to the
 width of the boundary region, $\sigma_t$, which depends on the wave packet size and the velocity variation, $\sigma_t=\frac{\sigma_s}{ \Delta V^2}$ . The derivation
 is given in \cite{ishikawa-oda-nakatsuka}
 
The momentum  distribution  is written as a sum of  two terms, 
\begin{eqnarray}
\frac{d  P}{ d^3{ K}_1  d^3{ K}_2}= 	
 	2 \left(K_1 \cdot K_2\right)^2  e^{    -{\sigma_s} ( \delta  P)^2}   \sum_{k=bulk, boundary} P_0^{k}    \left|G^{k}(\delta \omega)\right|^2, 
 	\label{total-distribution_{pi}}
\end{eqnarray}
where 
\begin{align}
 P_0^{k}
	&=
\begin{cases}
g^2 2^{-6} {(\sigma_{\pi})}^{-3/2} (E_{\pi} E_1 E_2)^{-1} C \tau_\pi \left(1-e^{-\frac{T}{\tau_\pi}}\right)	&\text{for bulk}, \label{norm}\\
g^2 2^{-6} {(\sigma_{\pi})}^{-3/2} (E_{\pi} E_1 E_2)^{-1} C \sqrt{ 2 \sigma_t}	&\text{for boundary},
\end{cases}
\end{align}
where $T=T_{\gamma}-T_{\pi}$,  $C$ is a constant  of energy dimension $E^1$ and  depends on the wave packet parameters. The   squares of  $ G({\delta \omega})$ in the asymptotic region is,
\begin{align}
|G({\delta\omega})|^2
=	e^{\sigma_t\over 4\tau_\pi^2}\times 
\begin{cases}
{e^{-{ ({T_0^R-T_\pi})^2\over \sigma_t}+{T_0^R-{T_\pi}\over {\tau_\pi}}}} \over {({T_0^R-{{T_\gamma}\over {\sigma_t}}}-{1\over 2 {\tau_{\pi}}})^2+({\delta\omega})^2} 	&\text{from boundary (${\sigma_t\over 2} {\delta\omega}^2\gg 1$)},\\
2\pi \sigma_te^{-\sigma_t ({\delta\omega})^2}		&	\text{from bulk (${\sigma_t\over 2} {\delta\omega}^2\ll1$)},
\end{cases}
\end{align}
where $T_0^R$ is the time that the wave packets intersect. The bulk term decreases rapidly with $\delta \omega $  and the boundary term  decreases slowly with an inverse power of the energy difference.

In the decay of the high energy pion of  $p_{\pi}=(E_{\pi},0,0, p_{\pi})$, the momenta of the final states are almost parallel to the pion. 
  In the boundary term,   $|G(\delta \omega )|^2$ decreases slowly at  $ K_i \rightarrow \infty$, and leads  a large contribution to the probability.

In the  transition,  the total energy is conserved  but  the kinetic energy is partly violated.  The  bulk contribution is narrow in the kinetic energy, and reveals  the golden rule. The  boundary contribution is  broad in the kinetic energy, and reveals the correction term. The deviation of the kinetic energy from the total energy 
is  the interaction energy 
$V_\text{int}= \langle \Psi| H_\text{int}| \Psi  \rangle$.   
The  coupling strength $g$  can be treated as  constant for the golden rule, where $k_{\gamma_i}\cdot k_{\gamma_{i'}} \ll m_q$.    However,  the boundary term is spread in  wide kinetic region of    $k_{\gamma_i}\cdot k_{\gamma_{i'}}$, which  includes a region   $k_{\gamma_i}\cdot k_{\gamma_{i'}} \gg m_q^2$. There,  this coupling strength    becomes 
 a function of $k_{\gamma_i}\cdot k_{\gamma_{i'}}$,  $g(  k_{\gamma_i}\cdot k_{\gamma_{i'}}) $, behaving as    
 \begin{eqnarray}
 g(  k_{\gamma_i}\cdot k_{\gamma_{i'}})=g\frac{m_q^2}{2k_{\gamma_i}\cdot k_{\gamma_{i'}}} 
 \end{eqnarray}
  \cite{liu}.  Here $m_q$ is the composite quark mass of a magnitude around  $\frac{m_p}{3}$ , where $m_p$ is the proton's mass.  Thus $P^{(d)}$ becomes maximum at around  $ k_{\gamma_i}\cdot k_{\gamma_{i'}} \approx  \frac{m_p^2}{2} $.  Its magnitude is proportional to 
the proton's mass. This behavior shows that the average interaction energy $\langle |H_{int}| \rangle$ is the order of the proton's rest energy, $m_p $. 


For a high energy pion, the initial and final waves overlap in wide area for  photons propagating in the parallel direction to  the pion. The boundary region becomes large in size , and gives large contribution to  the probability.   

\section{Positron annihilation }
Positron and   electron are described by the field $\psi(x)$, and photon   is by $A_{\mu}(x)$
in the Quantum Elecrodynamics, and   the interaction is  $e \bar \psi(x) \gamma_{\mu} \psi(x) A^{\mu}(x) $.   The para-positronium decay and the free positron annihilation  are  derived from  
 this interaction. The former one is also expressed by   an    effective interaction
 equivalent to the pion-two photon interaction. The latter one is described by 
the 2nd order perturbative expansion with respect to the above interaction.  $P(T)$ in these decays are studied.   

  \subsection{ Para-positronium decay }
 Para-positronium is even in the charge conjugation and  decays to two photons. The formula of decay probability  Eq.$	(\ref{total-distribution_{pi}})$ is applied after changing parameters with  suitable ones.  
   The average lifetime of the Para-positronium is much longer than that of the pion and the wave packet size is also longer.   The positronium decays and positron annihilation  in 
 porous material ,which are composed of small holes and many boundary regions, are analyzed. We will see that the boundary term  is enhanced.

  \subsection{Free positron annihilation }
 The annihilation amplitude  of the free positron and the free electron at rest    for those  of the central values of momentum and position,
\begin{eqnarray}
  \chi_{e_i}=(\vec p_{e_i},\vec X_{e_i}, \sigma_{e_i}), \chi_{\gamma_i}=((\vec p_{\gamma_i},\vec X_{\gamma_i}, \sigma_{\gamma_i})
  \end{eqnarray}
   for   the photons, the electron, and the positron, 
\begin{eqnarray}
\mathcal M=\langle {\chi}_{\gamma_1};{\chi}_{\gamma_2} |\int_0^T d x_1   \int_0^{t_1}   d x_2 H_{int}(x_1) H_{int}(x_2) |{\chi}_{e_1}, {\chi}_{{\bar e}_2} \rangle, \label{ positron_amplitude}
\end{eqnarray}
where $T$ is the time interval that the positron crosses a grain of the  target. The integrations over the coordinates ${\vec x}_i$, and  over the momentum ${\vec q}$ for the intermediate state are made using Gaussian integrations.
 
  The integration    over  times give the bulk and boundary terms, and 
 lead the amplitude to be  written as  Eq.$(\ref{total-amplitude_{pi}})$. 
 Substituting these,  we have the momentum  distribution   
\begin{align}
&\frac{1}{TL^3} \frac{d P}{ d^3{ k}_1  d^3{
 k}_2}
 	=   \frac{2}{m^2}(1+\frac{1}{4}(1-\cos \theta)+\frac{1}{2}(\frac{m}{E_{\gamma_1}}+\frac{m}{E_{\gamma_2}}))   \nonumber \\
 	 &[  e^{    -{\sigma_s} ( \delta  P)^2} ( P_0^{bulk}  \left|G_{bulk}(\delta \omega)\right|^2 +P_0^{b_t}   \left|G_{boundary_{(t)}}(\delta \omega)\right|^2) \nonumber \\
&+ P_0^{b_s}  e^{    -{\sigma_s} ( \delta   P)^2} \left|G_{boundary_{(s)}}(\delta \omega)\right|^2  ], \label{total-distribution_{positron}}
\end{align}
where 
Eqs.$(\ref{bulk})$ and $(\ref{boundary})$ 
are substituted, and 
\begin{align}
&  P_0^{bulk}    =   (E_{e^{+}} E_{\gamma_1} E_{\gamma_2})^{-1} C  	&\text{ bulk}, \label{norm}\\\
	& P_0^{b_{(t)}} =  (E_{e^{+}} E_{\gamma_1} E_{\gamma_2})^{-1} C \frac{\sqrt{ 2 \sigma_t}}{T}	&\text{ boundary~in ~time}, \\
	& P_0^{b_{(s)}}=(E_{e^{+}}  E_{\gamma_1} E_{\gamma_2})^{-1} C \frac{\sqrt{ 2 \sigma_t  2\sigma_s}}{TL} 	&\text{ boundary~in~space},
\end{align}
where $C$ is the constant \cite{ishikawa-oda-nakatsuka}. In silica powder, this size is  semi-microscopic of order few nano meter, and almost the same or slightly larger than  $ \sqrt{ \sigma_{\gamma}}$. In the present situation, the target is composed of silica particles of $L= 7$ nano meter,  and it is reasonable to assume  the ratios
$  \frac{ \sqrt{ 2 \sigma_s}}{L}$ and $ \frac{\sqrt{ 2 \sigma_t }}{T} $ are $ \frac{1}{10}-\frac{1}{100}$. 
The positron energy   is      $E_{e^{+}}=m_e $ with the energy uncertainty of 10 per cent.
      The spectrum 
  of the boundary term is of the universal form but its  magnitude has uncertainties due to the uncertainties on the wave packets.  This ambiguity could be studied by 
  a light scattering of the silica powder \cite{raman}.  

  
\section{Initial and final states  }
We apply  the decay probability Eq. $(\ref{total-distribution_{pi}})$ to  the  neutral pion in the process 
\begin{eqnarray}
e^{+}+e \rightarrow \phi \rightarrow \pi^{+} +\pi^{-}+ \pi^{0}
\end{eqnarray}, and 
that of the   positron  Eq.$(\ref{total-distribution_{positron}})$ in  the process
\begin{eqnarray} 
{}^{22}\text{Na} \rightarrow {}^{22} \text{Ne}^{*} +e^{+} +\nu, {}^{22}\text{N}e^{*} \rightarrow {}^{22}\text {Ne} + \gamma. 
\end{eqnarray}
 The former experiment is made in a high energy laboratory and the latter experiment is made  in a low-energy laboratory.

  \subsection{Wave packet shape  and size  }
 The total transition rate  $\Gamma$ derived from Eqs.  $(\ref{total-distribution_{pi}})$  and   $(\ref{total-distribution_{positron}})$ is independent of the wave packet parameters. This   is consistent with the general theorem given by Stodolsky \cite{Stodolsky}  \cite{Lipkin} \cite{Akhmedov}  \cite{Ishikawa-Tobita-ptp} on  stationary physical quantities. This theorem, however, is not applied to a non-stationary quantity such as $P^{(d)}$. In fact   $P^{(d)}$ derived from Eqs. $(\ref{total-distribution_{pi}})$  and   $(\ref{total-distribution_{positron}})$  depend on   the  forms 
 and sizes of the wave packets.   Up to here the Gaussian wave packet, which  decreases exponentially  in the position  and the momentum  
 and satisfies   the  minimum uncertainty  $ \delta x  \delta p ={\hbar} $, and   $\delta p=0$ for $\delta x =\infty$ is used. This
     is  idealistic for studying  the transition   for a finite time interval.   Other wave packet satisfying  $ \delta x  \delta p  \geq {\hbar} $  is shown to
      lead  almost equivalent results.  $\sigma_{\pi}$,  $\sigma_{\gamma}$, and  $\sigma_{\bar e}$ stand for $\sigma_s$ of  the pion, photon, and positron.  
      
     These particles interact with microscopic objects in matters and cause the final states to be produced, from which a number of the events and the probability 
      are determined.  Accordingly the packet parameters  in our formula are determined by these states in matter. This method has been shown valid 
      in \cite{ishikawa-shimomura-PTP, ishikawa-tobita-PTEP,ishikawa-tobita-ANP,ishikawa-tajima-tobita-PTEP}, and in   quantum 
      transition of two atoms in an energy transfer process in photosynthesis \cite{maeda-PTEP} .

\subsubsection{Sizes of wave functions :$\pi^{0}$ }
{ \bf  $\sigma_{\pi_0}$ }

In order for  the electron and the positron to produce a $\phi$ meson,   they are accelerated from average electron momentum  in matter, which is
 less than  $\frac{1}{10^{-10} \text{Meter}}$.  A relaxation time for these electron and positron in matter, beyond which  these lose  coherence due to environmental effect    is around $10^{-14}$ second, which corresponds to the   mean free path  $3 \times 10^{-6}$ Meters for the speed of light, and slightly shorter at  lower  energy.    $10^{-14}$ second and $10^{-7}$ Meters for  the spatial electron sizes   in matter are used.  The positron  is produced by the electron collision 
 with matter, and the length is the same as that of the electrons. During their  acceleration, the time interval that the  wave packets pass through at a spatial position is kept unchanged.     Although the  amplitude of  three pions, which is described by   the intermediate $\phi$ meson of   the Breit-Wigner form of the energy width of few MeV, peaks around the central energy, if the initial state has a fixed energy, each pion can have infinitesimal energy uncertainty.  Accordingly the width of the  $\phi$ meson  is related  neither  to the uncertainty of the pion's energy nor to the pion's wave packet size. Nevertheless,  the  above relaxation time  of the electron and positron results to an uncertainty of  the three pion's energy,    few meV. Thus the energy uncertainty of $\pi^0$ is governed by the relaxation time.  That leads   $3 \times 10^{-6}$ Meters for $\pi^0$ in the present process.    

   { \bf  $\sigma_{\gamma}$ in $\pi^0$ decay}

The detection process of the photon is governed  by its reaction with the atoms and the following coherent transitions by which electronic signals emerge  in the detector. They occur within finite spatial area occupied by the wavefunctions in solid.  The transition  amplitude  of the photon is described by the wave packet  of this  size. Thus $\sigma_{\gamma}$ represents    the spatial size of the electron wavefunction  in the configuration space  that the photon interacts with.  The initial process depends on the energy.   In the energy $0.5$ GeV,  majority of the events are the pair production due to nucleus electric field. Accordingly,    $\sigma_{\gamma}=\frac{ s_{\gamma}}{m_{\pi}^2}$, where $s_{\gamma}  \leq 1$ , and $s_{\gamma}=0.5$ is used for a following  estimation.
 
 $\sigma_{\gamma}$  and $T_i$ derived from $\sigma_i$ govern the magnitude of $ P^{(d)}$.  
 For high energy  colliding beam experiments, the sizes of the positron and the electron are determined by the spatial size of the electron wavefunction in matter.   $T_i=10^{-14}$ seconds from the relaxation time, and  $\tau=10^{-16}$ seconds.

 At the time interval, $T \gg \tau=\frac{1}{\Gamma}$,   the ratio    $T_i \Gamma$  becomes  $  \frac{T_i}{\tau}$. Now, $c \tau$ is $10^{-8}$ meter. Accordingly    the ratio   $\frac{T_i}{\tau} m_{\pi}^2 \sigma_{\gamma} \frac{1}{64}=\frac{10^{-14}}{10^{-16}} \times  0.5 \times \frac{1}{64}=0.8$. From this value   the probability  that one of the photon is in the energy range around the central energy ${E_{\gamma}}^0$,   $\frac{E_{\gamma}-{E_{\gamma}}^0}{{E_{\gamma}}^0} \leq 0.1$ is about 10 per cent. This would be consistent with the current uncertainty of the neutral pion's average lifetime.
\subsubsection{Sizes of wave functions :$\bar e $ }
{\bf  $\sigma_{e^{+}}$ }

   First we study the spatial size of  the positron wave packet  for a process  that the gamma from the positron annihilation is measured.    
 ${}^{22}$Na     is at rest and bound in matter.   
  The  spatial extension of ${}^{22}$Na's  wavefunction in the configuration space would be $\frac{1}{2000}$ of the electron wavefunction from the ratio of the masses.     The positron emitted from  ${}^{22}$Na  decay has this size in the direction perpendicular to its momentum, and that   in the parallel  direction can be much longer.  This    loses the  energy in matter in average $10^{-12}$ second  \cite{Rohrich- Carson} .  Hence  the  time interval in which  
  the positron wavefunction keeps the coherence or the average  relaxation time is $10^{-12}$ second.

   The wave packet size for  a detected  positron   is estimated  based on   the used detector.   When a  plastic
 scintillator  in which Benzen  is used,  the spatial size  of the Benzen molecule, around 1 nano meter,  shows the positron 
 wave packet size. 

{\bf  $\sigma_{Ps}$ }

Positronium are formed in porus material and  decays  there.   The size of the porus determines an effective size of the interaction 
area, and determines the time interval of the transition amplitude.

   { \bf $\sigma_{\gamma}$ in positron annihilation}

The dominant process of the  photon with matter   in the detector in this energy region, around a few hundreds KeV, is the photo-electric effect, in which the photon excites the atom.  The relevant 
spatial size is the size of atom, which is characterized by    $\pi \times (\frac{a_{\infty}}{2})^2$, where $a_B=\frac{1}{ m_e \alpha}$  is the Bohr radius.   
 Excited atoms make successive transitions and  produce many photons, electrons and ions of low energy.  These processes are expressed 
 by the time-dependent  Schroedinger equation  which  describe electrons, photons and ions. These states  are expressed by the wavefunctions of finite spatial extensions,  wave packets. The size of coherent area of these wave functions would be of   order few atomic sizes, due to decoherence caused by many atoms. 
 The wave packet size,  $\sigma_{\gamma}$ of the photon may be of a few atomic sizes.  The parameters may depend on the detector, \cite{Hossain,Moszynski}.  
   
  \subsection {  Boundary regions  }
The wave functions of the electron and positron overlap at the boundary region of the matter, and their annihilation takes place.   
  The area is  large, and the events increases in porus material.    The porus size  determines an effective size of the area  and the time interval of the transition.   The transition amplitudes and probabilities depend on these sizes.   
  That is used in the positron experiments.

 For experiments  that use  small powders,   electrons are   inside of  the small region, and  the interaction takes place  in the inside or at the boundary region. 
 The transition amplitudes and probabilities depend on these sizes.  
 \subsection{Energy resolution}
 An idealistic  detector that detects and gives an energy  of a particle or a wave directly does not exist. For its  measurement,  signals caused by its reactions 
 with matter are read first and is converted to the energy using a  conversion rule justified by other processes.  The energy is measured within finite uncertainty.
 This is the energy resolution, and all the detector have the finite energy resolution. This causes an  experimental uncertainty. 
     The energy resolution, $\sigma(E)$,   has  various origins such as   a statistical one  and an  intrinsic one. 
  That is written as 
 \begin{eqnarray}
\sigma(E)=\sigma_{statitics}(E)+\sigma_{intrinsic}(E),
 \end{eqnarray}
 where  $ \sigma_{statitics}(E)$  is determined normally from  Poisson statistics and other is written as  $ \sigma_{intrinsic}(E) $,  in which  an effect 
 due to the finite size of wavefunctions, Eq.$(\ref{total-probability_{pi}} )$, is included. The former  depends on the detector's type, and the latter 
does not and has universal properties  regardless of 
 detector type. In scintillation  detector, an electric signal of a $\gamma$-ray  is obtained according 
to the number of the scintillation photons $N$, and the energy resolution,  $\sigma_{statitics}(E)$, is given by  
\begin{eqnarray}
\sigma_{statitics}(E)= 2.35 \sqrt \frac{  F}{  N} E,
\end{eqnarray}
where $N$ is a number of the sample and $F$ is a correction factor, the Fano-factor. For NaI(TI),   $F=1$, and  $ \sigma_{statistics}(E)/{\langle E \rangle}$ is around $5 -10$ per cent , and  
 the energy resolution is $25-50$ keV  for the energy $500$ keV. Ge detector is of different mechanism of much smaller statistical uncertainty, due to the small $F$ and 
 large $N$. The distribution around the central value decreases exponentially with $E$.  

  The wave-packet  size determined by  the size of the atom is  $\pi(10^{-10})^2$ M$^2$ and 
   should be almost  the same in NaI(TI) and Ge detectors, and leads  to the energy uncertainty, $\sigma_{intrinsic}(E) =1$ keV.     Accordingly  in the NaI, $\sigma_{statitics}(E)$ is the dominant one and $\sigma_{intrinsic}(E)$ is negligible,
   but in Ge detector, $\sigma_{intrinsic}(E) $ shares the substantial part.
\subsection{Energy distribution} 
 The energy distributions  of the bulk term and the boundary term are very different. That  from $\Gamma$  for the plane waves under ASI  is proportional to   $\delta(E_i-E_f)$,  
   but for the wave packets that behaves as   $e^{-(\frac{ \delta \omega }{\sigma (E)})^2}$, where the width  is of universal nature and behaves differently 
    from   those of  statistical one.   That of  $P^{(d)}$ decreases in $E^{-n}$, where $n \geq  0$ depends on the decay dynamics.  
 $P^{(d)}$ can be identified easily in the energy region $E \gg \sigma(E)$ if the relative fraction
 over  $\Gamma T $ is of substantial 
magnitude of the order ${10^{-3}}$ or larger, even with the detector of large energy resolution.  
Despite of  large energy resolution,  NaI(Tl)  scintillator 
is useful for the confirmation of $P^{(d)}$. The detector  of much smaller resolution such as the Ge detector  is also useful.

\section{Experimental confirmations }

As  $P^{(d)}$ possesses  many unusual properties,   phenomena originated from  $P^{(d)}$  reveal intriguing  properties.   By detecting  these events,   $P^{(d)}$ can be confirmed.  $\Gamma$ has been well established, and phenomena of $\Gamma $ origin 
 have been understood precisely  with a help of  numerical methods.  They are compared with  the data from 
the natural phenomena and  observations. 
If clear disagreements are   found,    and if it is   resolved by  $P^{(d)}$,  this may 
confirm  $P^{(d)}$.    
\subsection{Magnitude of $P^{(d)}$}
A magnitude of $P^{(d)}$ for para-positronium decay, $P^{(d)}(pp)$, and direct annihilation, $P^{(d)}(da)$,  is estimated and  given in Figure. They  depend on the size and shape of the wave packets. We use the value $\sigma_{\gamma} > 10^{-20} {\text m}^2$, and the Gaussian wave function and power law wave function, and find    
\begin{eqnarray}
P^{(d)}(pp)&=&  10^{-12}~; \text{Gaussian WF} \nonumber \\
&=& 3 \times 10^{-4} ~\text{Power law WF} \nonumber \\ 
P^{(d)}(da)&=& 2\times 10^{-6}~; \text{Gaussian WF} \nonumber \\
&=&  3\times 10^{-4}~; \text{Power law  WF}. \nonumber  
\end{eqnarray}
At the moment we are not aware of the precise shape and size of the wave function. Light scattering may be useful for a study of the wave function \cite{raman}. 
    
The photon distribution is modified by  $P^{(d)}$ in  the positron annihilation and positronium decay.  The  high energy side is not affected by the modified energy by the Compton scatterings, which is not true on the low energy side.   
By measuring multiple  coincident photons in the high energy regions, clear signals may be obtained.   Although accidental coincident events may contribute,
  the separation of  them can be   made and   
and  events  of $P^{(d)}$ origin in the  data is estimated.       It is our expectation that with $10^8$ events of the positron annihilation a confirmation of $P^d$ could be in scope. 
\begin{figure}[t]
\centering{\includegraphics[scale=.8,angle=0]{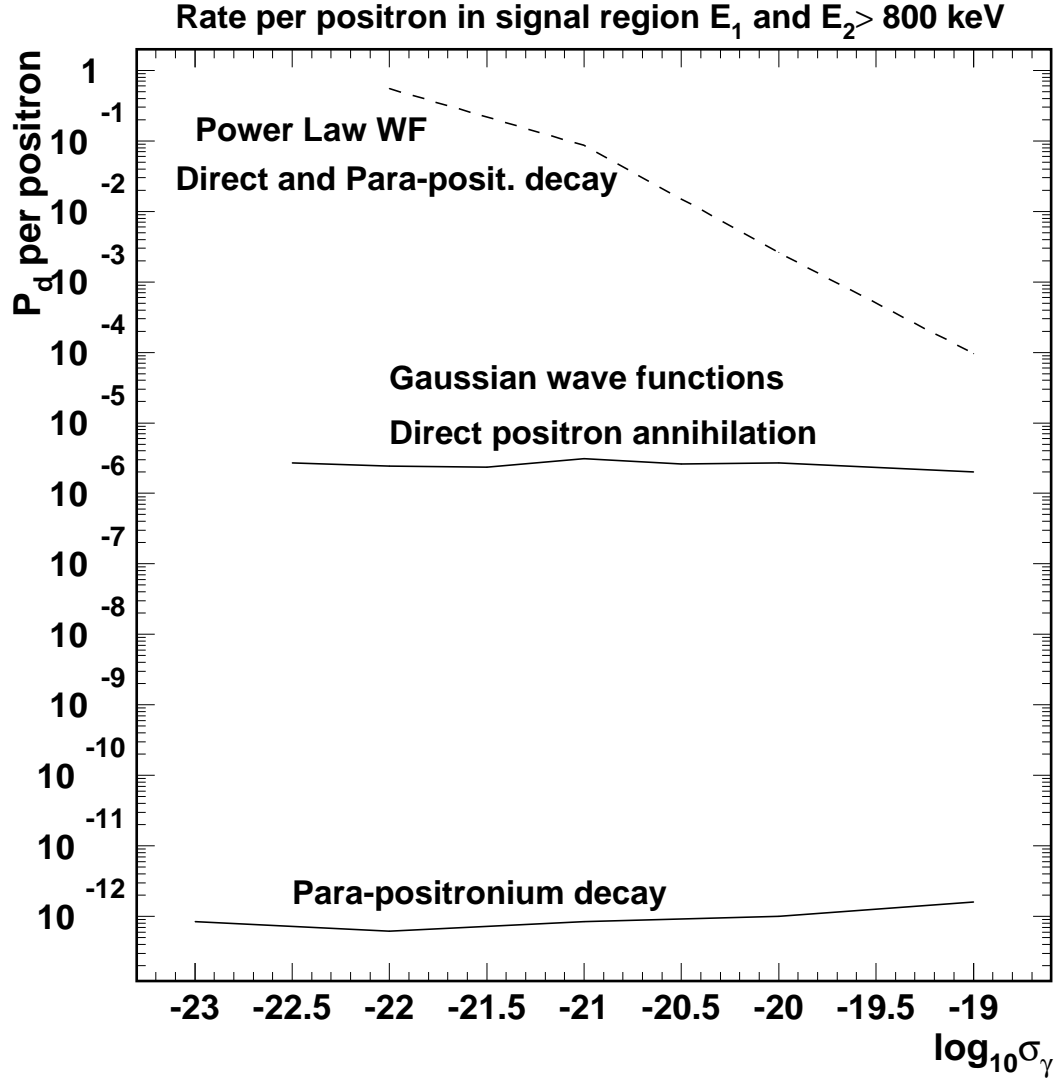}
\caption{The variation of the expected ratio of events per stopping 
positron in silica with $\sigma_\gamma$ for different assumed 
situations. The upper curve shows the ratio if direct wave functions 
are dominant. The middle curve shows the ratio if direct annihilation 
dominates and the lower curve if the annihilation is dominantly through 
the decay of para-positronium.}} 
\end{figure}

GEANT4\cite{GEANT4}is a simulation program that includes the transition probability and the detector performance. The probabilities  derived from the 
golden rule are employed . Hence this    is quite useful for  analyzing the natural phenomena including   the detector's response and backgrounds. 
  Comparing the events derived from the golden rule of the standard theory with the observations, we are 
able to see if a non-standard component is included.  

\subsection{  Backgrounds from decay (  annihilation ) in flight}
  The signals from the decay or the annihilation in flight are in  energy regions  different from those at rest and give background. 
  Positron loses its energy  in insulator  in pico seconds \cite{stopping-power}, and stops.  A photon produced before the stop  has an energy  higher than
     $ m_e $ and its contribution is estimated in two steps.   
     
    The average positron lifetime, due to the annihilation or the decay  is,   100-500 pico seconds, which  depends on 
  various conditions. Hereafter we use 200 pico seconds for the average life  time and 2 pico second for the thermalization time.   The annihilation events of the positron in flight over that at rest is  less than 
   the ratio ,   $\frac{2}{200}=10^{-2}$. The experimental value seems to be less than $10^{-3}$ or $10^{-4}$  \cite{inflight-annihilation}.  
       Among the events of  energy      $E_1+E_2 > 2m_e c^2$, a fraction  in the energy region $E_1+E_2-2 m_e c^2 \geq 3\sigma E$, where $\sigma E$  is
     the width of NaI(Tl) detector, is obtained as $\frac{1}{230}$  from  Bethe's formula  \cite{bethe}.  A further suppression factor $\frac{1}{10}$ is multiplied due to a specific  configuration of the detector setup of the present experiment.  Combining these  numbers, the fraction is $ 0.43 \times 10^{-6}$ or $ 0.43 \times 10^{-7} $ . 
   This gives the magnitude of the background from the inflight annihilation, is less than $  10^{-6}$.

\subsection{Uncertainties}
Possible  sources of uncertainties and   ambiguities are matter effects, accidental coincident events  (double hits) , and   environmental gammas.

   The photon spectrum in the high energy region 
is not modified  by  Moeller scattering,    photo-electric effect, the Compton effect,  and the pair
 production. Accordingly  the matter effects are irrelevant.    The   environmental gammas  or those of cosmic ray origins   are avoided by selecting 
  coincedent events of multiple gammas.  In two gamma's case, the coincedence between   one gamma from Ne radiative decay and  another from the positron
   annihilation are taken. In  three gammas case, the coincedence between   one gamma from${}^{21}$Ne$^{*}$ radiative decay and two photons   from the positron
   annihilation are taken.  In these multiple coincident events, there remain accidental coincident events  (double hits).   Because their strength depends upon 
   the initial positron flux  and the spectrum has different momentum dependence than the signal from $P^{(d)}$, it is possible to disentangle them following
    Appendix B.

 \subsection{ Related processes}
Para-positronium decays is included in the text.  Other spin component, Orth-Positronium ,  may be used for $P^d \neq 0$ test. However,   Orth-Positronium has much longer life-time and $P^{(d)}$  becomes much smaller. Its effect is difficult to observe experimentally.     
  $P^{(d)}$ in  nuclear's gamma and beta decays become also sizable, and can be non-vanishing even in  processes of $\Gamma=0$. Various   selection rules
  are   valid only to $\Gamma$, but to $P^{(d)}$.    A role of $P^{(d)}$  is important then.

\section{Summary and prospects}
$P^{(d)}$ would be confirmed from the photon's distributions experimentally.  
   
  (1) The energies of the photons  in the positron annihilation at rest from the golden rule satisfy  
$E_{\gamma_1}+E_{\gamma_2}=2m_e $, whereas  those  from $P^{(d)}$ satisfy  
$E_{\gamma_1}+E_{\gamma_2} < 2m_e $ or $E_{\gamma_1}+E_{\gamma_2} >2m_e $.  The photon loses its energy by the Compton scattering, and that produced by the  golden rule can  be detected  in the former region, but not   in the latter region.   The events of the energies    $E_{\gamma_1}+E_{\gamma_2} >  m_e $  are generated only by   $P^{(d)}$, and may be worthwhile  for its  confirmation.   

(2) For the neutral pion,  our finding  $ P^{(d)} \approx O(0.8)$  suggest that    for the  analysis   $ P^{(d)}$ must be implemented.
  The previous  large    uncertainty of about  $10$ per cent in the life time would be due to $P^{(d)}$, and will be reduced  in an analysis that includes  
   $P^{(d)}$.

(3) Tagging $\pi^{+}$ and $\pi^{-}$ in the process $e+\bar e \rightarrow \phi, \phi \rightarrow \pi^{+} +\pi^{-}+\pi^0$, the $\pi^{0}$ momentum is determined, and the photon spectrum is computed.   Due to $P^{(d)}$, this spectrum deviates from  the golden rule. If the deviation  is  observed,   $P^{(d)}$ will be confirmed.    

(4) Many-body wave functions of  $ \delta E= E_{initial}-(E_{\gamma_1}+E_{\gamma_2}) \neq 0$
 have  interaction energies,  which are  independent of the frequency of each wave.  This leads  an extra component to the energy momentum tensor in addition to  those   proportional to the frequencies.   Normal detection processes measure the wave's frequencies, but these interaction energies.   Accordingly,  this corresponds to an invisible energy. This state may  be considered as  a kind of  halo.     
 
 (5) Once the confirmation of $P^d$ is made, (a)  methods to reduce  current uncertainties in the experiments and  (b) mechanisms to solve current 
puzzling phenomena will be found.

\subsection*{Acknowledgments}
 This work was partially supported by a 
Grant-in-Aid for Scientific Research ( Grant No. 24340043). The authors  thank  Dr. K. Hayasaka, Dr. K. Oda,
and Mr. H. Nakatsuka
for useful discussions.

\appendix
\section*{ Appendix }
\section{Free positron annihilation}
\subsection{Amplitude}
An  amplitude for a free positron annihilation is 
\begin{eqnarray}
& &M=\int_0^T dt_1\int d^3 x_1 \int_0^{t_1} d t_2 d^3 x_2\langle \gamma_1,\gamma_2|H_{int}(x_1) H_{int}(x_2)| e, \bar e \rangle,\\
& &=\frac{1}{2}\int_0^T dt_1\int_0^T d t_2 d^3x_1 d^3x_2  \langle \gamma_1,\gamma_2|T(H_{int}(x_1) H_{int}(x_2))| e, \bar e \rangle, 
\end{eqnarray}  
where $H_{int}(x)$ is the interaction part of QED and the initial and final states are wave packets, and 
\begin{eqnarray}
T(H_{int}(t_1)H_{int}(t_2))=\theta(t_1-t_2) H_{int}(t_1)H_{int}(t_2)+\theta(t_2-t_1) H_{int}(t_2)H_{int}(t_1).
\end{eqnarray} 
 Applying the Wick's theorem,
\begin{eqnarray}
& &M=\sum : : \gamma_{\mu}S_F(x_1-x_2) \gamma_{\nu} + \cdots \\
& &=\bar u(p_1)[ \gamma \epsilon(k_1) \frac{\gamma (p_1-k_1)+m_e}{(p_1-k_1)^2-m_e^2} \gamma \epsilon(k_2)+ \gamma \epsilon(k_2) \frac{\gamma( p_1-k_2)+m_e}{(p_1-k_2)^2-m_e^2} \gamma \epsilon(k_1)] v(p_2) \nonumber\\
& &=-\bar u(p_1)[ \frac{\gamma k_1 \gamma \epsilon(k_1) \epsilon(k_2)+2 \epsilon(k_1)p_1 \gamma \epsilon(k_2)}{ 2p_1k_1} +\frac{ \gamma k_2 \gamma \epsilon(k_2) \gamma \epsilon(k_1)+2\epsilon(k_2) p_1 \gamma \epsilon(k_1)}{2p_1k_2} ] v(p_2) \nonumber\\
\end{eqnarray}
where  
\begin{eqnarray}
\bar u(p_1) \gamma \epsilon(k_1) ( \gamma p_1+m_e)=2\epsilon(k_1)p_1\bar u(p_1)
\end{eqnarray}
and the similar one for the $v(p_2)$ were substituted.For  $p_1=(m,o), p_2=(m,o)$
it follows 
\begin{eqnarray}
\sum_{spin} |M|^2=\frac{8}{m_e^2}[ 1+\frac{1}{4}(1-\cos \theta )+\frac{1}{2}(\frac{m}{k_1^0}+\frac{m}{k_2^0})].
\end{eqnarray}
Note that this is slightly different from that of the positronium  decays.

\subsection{Boundary in space and time}
\subsubsection{ Amplitude}
In scatterings in laboratory flame   where the target is composed of small particles of the volume $L^3$, the momentum dependent 
amplitudes in   
the bulk and boundary terms, of Eq.(15) in  August 25 version are replaced with 
\begin{eqnarray}
& &M_{bulk}=\sqrt{2 \pi \sigma_t} e^{-\frac{\sigma_t}{2} ( \delta \omega )^2} ({2 \pi \sigma_s})^{3/2} e^{-\frac{\sigma_s}{2} ( \delta {\vec p} )^2} \theta({\vec X_{\gamma}},volume)\\ 
& &M_{boundary_{(t)}}=\sqrt{\frac{2  \sigma_t}{\pi}}  \frac{1}{ -i \sqrt {\sigma_t } \delta \omega +1} 
({2 \pi \sigma_s})^{3/2} e^{-\frac{\sigma_s}{2} ( \delta {\vec p} )^2} \theta({\vec X_{\gamma}},b_t) \nonumber \\
& & \\ 
& &M_{boundary_{(s)}}=\sqrt{\frac{2  \sigma_t}{\pi}} \frac{1}{ -i \sqrt{ \sigma_t} \delta \omega +1}   ({2 \pi \sigma_s})^{2/2}\sqrt{\frac{2 \sigma_s}{\pi}} \times \theta({\vec X_{\gamma}},b_{(t,s)}) \nonumber \\
& &[\frac{1}{ -i \sqrt{ \sigma_s} \delta p_z +1} e^{-\frac{\sigma_s}{2} ( (\delta {\vec p_x} )^2 + ( \delta {\vec p_y} )^2)}+(z ,x,y) \rightarrow (x,y,z ),(y,z,x)  ]. \nonumber \\
& & 
\end{eqnarray}
where $\theta({\vec X_{\gamma}},volume)$, $\theta({\vec X_{\gamma}},b_t)  $, and $ \theta({\vec X_{\gamma}},b_{(t,s)}) $ show that the intersection
 of trajectories are  in the inside of the volume $L^3$,    in the boundary in time,  and  in the boundary in space  and time.   
 
 The momentum dependent term in the bulk,  Eq.(21), and the boundary term in time, Eq.(22), lead the probability of the same form as before,  
 \begin{eqnarray}
& & |G_{bulk}(\delta \omega )|^2 =(\sqrt{2 \pi \sigma_t} e^{-\frac{\sigma_t}{2} ( \delta \omega )^2})^2  ({2 \pi \sigma_s})^{3}  \label{bulk}\\
& & |G_{boundary_{(t)}}(\delta \omega )|^2 = |\sqrt{\frac{2  \sigma_t}{\pi}} \frac{1}{ -i \sqrt \sigma_t \delta \omega +1}|^2
( {2 \pi \sigma_s})^{3}  \nonumber 
\end{eqnarray}
 but the space-boundary term  
  \begin{eqnarray}
& &e^{-\sigma_s( \delta {\vec p})^2} |G_{boundary_{(s)}}(\delta \omega )|^2 =
 (\sqrt{\frac{2  \sigma_t}{\pi}} \frac{1}{| -i \sqrt \sigma_t \delta \omega +1|} \sqrt{\frac{2  \sigma_s}{\pi}}(2 \pi \sigma_s))^2  \label{boundary}\nonumber \\
& & \times [|{\frac{1}{ -i \sqrt \sigma_s \delta p_z +1}}|^2   e^{-{\sigma_s} ( (\delta {\vec p_x} )^2 + ( \delta {\vec p_y} )^2)}+|(x,y,z )|^2+|(y,z,x)|^2  ]. 
\end{eqnarray}
is different. The momentum dependence of the bulk term and that of the time-boundary are spherically symmetric  as  before but that of 
the space-boundary is  asymmetric. 

\subsection{ Normalization of Probability: summation over the positions}
The integration over the positions ${\vec  X}_{\gamma_l}$, and over the  position ${\vec  X}_{e^{+}}$  in the region of $L^3$ and the time interval $T$, and  for the boundary of the width $ \sqrt{ 2\sigma_t}$ and $ \sqrt{ 2 \sigma_s}$ are, 
\begin{eqnarray}
& &\int  d {\vec X_{e^{+}}} \int \frac{d {\vec X_{\gamma_1}} d {\vec X_{\gamma_2}}}{(2 \pi)^6} e^{-2R({\vec X_{\gamma_i}})} \theta({\vec X}_{\gamma},volume )=T L^3 \\
& &\int  d {\vec X_{e^{+}}}\int \frac{d {\vec X_{\gamma_1}} d {\vec X_{\gamma_2}}}{ (2 \pi)^6} e^{-2R({\vec X_{\gamma_i}})} \theta({\vec X}_{\gamma},b_t ) =\frac{\sqrt {2\sigma_t}}{T} (T  L^3) \\
& & \int  d {\vec X_{e^{+}}} \int \frac{\int d {\vec X_{\gamma_1}} d {\vec X_{\gamma_2}}}{(2 \pi)^6} e^{-2R({\vec X_{\gamma_i}})} \theta({\vec X}_{\gamma},b_{s,t})=\frac{\sqrt{2 \sigma_t  2 \sigma_s}}{TL} (T L^3) \nonumber \\
& & .
\end{eqnarray}
Substituting these ,  we have the momentum  distribution   
\begin{align}
&\frac{1}{TL^3} \frac{d P}{d^3{ k}_1 d^3{
 k}_2}
 	=   \frac{2}{m^2}(1+\frac{1}{4}(1-\cos \theta)+\frac{1}{2}(\frac{m}{E_{\gamma_1}}+\frac{m}{E_{\gamma_2}}))   \nonumber \\
 	 &[  e^{    -{\sigma_s} ( \delta  P)^2} ( P_0^{bulk}  \left|G_{bulk}(\delta \omega)\right|^2 +P_0^{b_t}   \left|G_{boundary_{(t)}}(\delta \omega)\right|^2) \nonumber \\
&+ P_0^{b_s}  e^{    -{\sigma_s} ( \delta  P)^2} \left|G_{boundary_{(s)}}(\delta \omega)\right|^2  ], \label{total-distribution_ap}
\end{align}
where 
Eqs.$(\ref{bulk})$ and $(\ref{boundary})$ 
are substituted, and 
\begin{align}
&  P_0^{bulk}    =   (E_{e^{+}} E_{\gamma_1} E_{\gamma_2})^{-1} C  	&\text{ bulk}, \label{norm}\\\
	& P_0^{b_{(t)}} =  (E_{e^{+}} E_{\gamma_1} E_{\gamma_2})^{-1} C \frac{\sqrt{ 2 \sigma_t}}{T}	&\text{ boundary~in ~time}, \\
	& P_0^{b_{(s)}}=(E_{e^{+}}  E_{\gamma_1} E_{\gamma_2})^{-1} C \frac{\sqrt{ 2 \sigma_t  2\sigma_s}}{TL} 	&\text{ boundary~in~space},
\end{align}
where $C$ is the constant. In the present situation, the target is composed of silica particles of $L= 7$ nano meter,  and it is reasonable to assume
  $E_{e^{+}}=m_e (1 \pm \frac{1}{10})$,  $  \frac{ \sqrt{ 2 \sigma_s}}{L}, \frac{\sqrt{ 2 \sigma_t }}{T} \approx \frac{1}{10}-\frac{1}{100}$. The spectrum 
  of the boundary term is of the universal form but its  magnitude has uncertainties due to the uncertainties on the wave packets.  This ambiguity could be studied by 
  a light scattering of the silica powder.  
 
\subsection{Non-Gaussian wave packet}
Function $e^{-\mu r }$, where $r=|{\vec x}|$ and $\mu $ is a constant, decreases rapidly at large distance $r$ but has a singularity at $r=0$. Its Fourier transform is $\frac{1}{(p^2+\mu^2)^2}$, and  is decreasing slowly in the momentum.  Accordingly, the wave packet of this form leads a probability different from the Gaussian one.   This  is studied hereafter.
\subsubsection{ Amplitude}
For the non-Gaussian wave packets, the momentum dependent 
amplitudes in   
the bulk and boundary terms  are 
\begin{eqnarray}
& &M_{bulk}=       f_0 \frac{ \frac{2}{t_0} }{ \omega^2+ (\frac{1}{t_0})^2} \times f_0  {8 \pi}  \frac{1}{r_0}    \frac{1}{(  (\frac{1}{r_0})^2 + ({\delta {\vec p}})^2)^2}              \theta({\vec X_{\gamma}},volume) \\
& &M_{boundary_{(t)}}=f_0      \frac{1}{i \omega + \frac{1}{t_0} } \times f_0  {8 \pi}  \frac{1}{r_0}    \frac{1}{(  (\frac{1}{r_0})^2 + (\delta {\vec p})^2)^2} \theta({\vec X_{\gamma}},b_t) ,
\end{eqnarray}
where $f_0= \frac{1}{ \sqrt{\pi r_0^3}}$, and $t_0$ and $ r_0$ are determined from the size of the Coulomb wave function and for NaI are given at the end of this Appendix.    These  lead the probability of the same form as before,  
 \begin{eqnarray}
& & |G_{bulk}(\delta \omega, \delta {\vec p} )|^2  =[ f_0 \frac{ \frac{2}{t_0} }{ \omega^2+ (\frac{1}{t_0})^2} \times f_0  {8 \pi}  \frac{1}{r_0}    \frac{1}{(  (\frac{1}{r_0})^2 + ({\delta {\vec p}})^2)^2} ]^2   ,                                          \label{bulk}\\
& & |G_{boundary_{(t)}}(\delta \omega, \delta {\vec p} )|^2 = [f_0      \frac{1}{i \omega + \frac{1}{t_0} } \times f_0  {8 \pi}  \frac{1}{r_0}    \frac{1}{(  (\frac{1}{r_0})^2 + (\delta {\vec p})^2)^2}  ]^2 .  \nonumber 
\end{eqnarray}
  The integration over the positions ${\vec  X}_{\gamma_l}$, and over the  position ${\vec  X}_{e^{+}}$  are also the same as before.
We have the momentum  distribution   
\begin{align}
&\frac{1}{TL^3} \frac{d P}{d^3{ k}_1 d^3{
 k}_2}
 	=   \frac{2}{m^2}(1+\frac{1}{4}(1-\cos \theta)+\frac{1}{2}(\frac{m}{E_{\gamma_1}}+\frac{m}{E_{\gamma_2}}))   \nonumber \\
 	 &\times [    ( P_0^{bulk}  \left|G_{bulk}(\delta \omega, \delta {\vec p}))\right|^2 +P_0^{b_t}   \left|G_{boundary_{(t)}}(\delta \omega, \delta {\vec p})\right|^2)] 
\end{align}
where 
Eqs.$(\ref{bulk})$ and $(\ref{boundary})$ 
are substituted, and 
\begin{align}
&  P_0^{bulk}    =   (E_{e^{+}} E_{\gamma_1} E_{\gamma_2})^{-1} C_{Coul}  	&\text{ bulk}, \label{norm}\\\
	& P_0^{b_{(t)}} =  (E_{e^{+}} E_{\gamma_1} E_{\gamma_2})^{-1} C_{Coul} \frac{\sqrt{ 2 \sigma_t}}{T}	&\text{ boundary}, \nonumber 
\end{align}
where $C_{Coul}$ is a constant which is related with $C$, and  $r_0=\frac{\sqrt{\sigma_s}}{50}$ and  $t_0= \frac{\sqrt{\sigma_t}}{50}$. We leave $C_{Cou}$ as 
a parameter  for a while. 
\section{Duplicate (accidentally coincident and pile-up) events}
Suppose the probability is a sum of duplicate (accidental coincident and pile-up ) events and $P^{(d)}$ events
\begin{eqnarray}
f(p_1,p_2)= c_0 g(p_1) g(p_2)+g(p_1,p_2)
\end{eqnarray}
where $g(p_1)$ and $g(p_1,p_2)$ are known theoretically, but $c_0$ in experiments is unknown. Define
 an error  function 
\begin{eqnarray}
& &I_{error}(c)= \int d p_1 d p_2 (f_{exp}(p_1,p_2) -c g(p_1) g(p_2))^2 \\
& &=\int d p_1 d p_2[ (c-c_0)^2 (g(p_1) g(p_2))^2+(g(p_1,p_2))^2-2 (c-c_0)g(p_1) g(p_1,p_2)g(p_2)] \nonumber \\
& &= (c-c_0)^2 A_2 -2(c-c_0) A_1+A_0 =A_2(c-\tilde c_0)^2+A_0  -A_1^2/{A_2}   \nonumber 
\end{eqnarray}
\begin{eqnarray}
A_2=\int d p_1 d p_2 (g(p_1) g(p_2))^2,A_1= \int d p_1 d p_2 g(p_1) g(p_1,p_2)g(p_2),A_0=
\int d p_1 d p_2 (g(p_1,p_2))^2
\end{eqnarray}
Plot $ I_{error}(c)$ as a function of $c$ and obtain the minimum value $D=A_0-A_1^2/{A_2}$,
\begin{eqnarray}
D=-[\int d p_1 d p_2 g(p_1) g(p_1,p_2)g(p_2)]^2/[\int dp_1 dp_2 (g(p_1) g(p_2))^2]+ \int dp_1 dp_2 g(p_1,p_2)^2 
\end{eqnarray}
 $P^{(d)} =0$,  $g(p_1,p_2) =0, and~  D =0$,  for $P^{(d)} \neq 0$,  $g(p_1,p_2) >0, and~  D >0$.
Experimental determination of $D >0$ may be feasible.


\begin{thebibliography}{}
\bibitem{raman} Raman scattering experiment has seen anomalous signals. Private communications from Professor.M. Takesada.
\end{thebibliography}
\end{document}